\documentclass[12pt]{article}
\usepackage{amsmath,amssymb,graphicx}
\usepackage{caption}
\usepackage{authblk}
\usepackage[a4paper, margin=2.5cm]{geometry}
\usepackage[numbers,square]{natbib}
\usepackage{subcaption}
\usepackage[title]{appendix}
\usepackage[T1]{fontenc}
\usepackage[utf8]{inputenc}
\usepackage{xcolor}
\usepackage[normalem]{ulem}
\usepackage{orcidlink}
\usepackage{placeins}

\usepackage{hyperref} % پکیج برای لینک‌های قابل کلیک
\hypersetup{
    colorlinks=true, 
    linkcolor=blue, 
    citecolor=blue, 
    urlcolor=blue 
}

\title{\textbf{Poloidal Field Amplification through Compression–Shear Dynamics in Schwarzschild Accretion: Pathways to MAD States}}

\author[1]{Malihe Mousapour Gharghabi\,\orcidlink{0009-0000-5748-8316}
 \thanks{Corresponding author: m.mousapour@khayyam.ac.ir}}
 
\author[1,2]{Jamshid Ghanbari}
\author[3]{Mahboobe Moeen Moghaddas}

\affil[1]{Department of Sciences, Khayyam University of Mashhad}
\affil[2]{Faculty of Basic Sciences, Ferdowsi University of Mashhad}
\affil[3]{Department of Sciences, Kosar University of Bojnord}

\date{}

\begin{document}

\maketitle

\begin{abstract}
The amplification of magnetic fields in black hole accretion flows governs key high-energy phenomena such as magnetically arrested disks and relativistic jets. We develop a semi-analytical general relativistic framework that extends classical compressional amplification models by incorporating rotational shear, and apply it to large-scale poloidal magnetic field evolution in accretion flows around a Schwarzschild black hole. By parameterizing the azimuthal velocity as a fraction of the Keplerian value ($\xi \in [0,1]$), from purely radial infall ($\xi=0$) to Keplerian rotation ($\xi=1$), we examine the combined effects of radial compression and shear. Purely radial flows maximize amplification of both $_r B$ and $_\theta B$ due to strong compression. In rotating flows, a distinct dichotomy emerges: sub-Keplerian regimes ($\xi<1$) preferentially enhance $_r B$, whereas Keplerian rotation strengthens $_\theta B$ via shear. The transition from subsonic outer regions to supersonic relativistic inner regions further accelerates magnetic growth, revealing effects absent in earlier analytical treatments. These results show that rotational support controls both amplification efficiency and magnetic geometry, with sub-Keplerian phases particularly favorable for advecting the radial flux required for MAD formation. This work provides an analytical bridge between classical accretion theory and modern GRMHD simulations, with implications for X-ray binaries, AGNs, and EHT-scale systems.
\end{abstract}


\clearpage

\section{Introduction}\label{sec:introduction}

Astrophysical accretion disks are among the most luminous and dynamic structures in the universe, playing a pivotal role in phenomena ranging from star and planet formation to the high-energy emissions of X-ray binaries (XRBs) and active galactic nuclei (AGNs) \citep{frank2002accretion}. 
In accreting black hole systems, gravitational potential energy---and, in rapidly rotating cases, spin energy---can be converted into radiation with high efficiency, as captured by the standard thin-disk framework \citep{Shakura1973,Novikov1973}. At accretion rates corresponding to luminosities between $\sim10^{-2}$ and $\sim0.5$ of the Eddington limit, the flow remains in a geometrically thin, radiatively efficient regime, whereas departures from this range lead to radiatively inefficient solutions such as advection-dominated accretion flows (ADAFs) \citep{Narayan1995,Abramowicz1995}. Although the classical model of \citep{Novikov1973} assumes that emission terminates at the innermost stable circular orbit (ISCO), magnetohydrodynamic (MHD) effects are now known to enhance energy release near the ISCO and alter the overall radiative efficiency \citep{Gammie1999,Krolik1999}.
In XRBs, disks around stellar-mass black holes (BHs) power intense X-ray radiation, while in AGNs, disks feeding supermassive BHs at galactic centers generate immense luminosities and launch relativistic jets that inject energy into their environments, regulating galaxy formation and evolution via feedback mechanisms \citep{2023ApJ...944..182D, 2021NewAR..9201594C}.

The presence of magnetic fields is, in fact, fundamental to both the dynamics and the observability of black hole accretion. In the absence of magnetic fields, for instance, spherically symmetric accretion onto a BH has an extremely low radiative efficiency of $\eta \sim 10^{-8}$, making such objects effectively invisible. In sharp contrast, even an initially irregular magnetic field can substantially enhance the radiative output, reaching luminosities of up to $\sim 0.1,\dot{M}c^2$ \citep{BisnovatyiKogan1974,BisnovatyiKogan1976}. This occurs because flux freezing during gravitational infall amplifies the field, leading to equipartition between kinetic and magnetic energies and the formation of quasi-stationary flows near the black hole.
Magnetic fields are therefore widely regarded as key agents in accretion systems, enabling angular momentum transport, jet launching, and disk–jet coupling. A long-standing challenge in accretion disk theory is to understand the mechanism of angular momentum transport that allows matter to spiral inward. While small-scale turbulence driven by the magnetorotational instability (MRI) is now recognized as the primary source of effective viscosity \citep{1995ApJ...446..741B}, the launching of powerful collimated jets requires large-scale ordered magnetic fields, the origin and evolution of which remain active areas of investigation.

The generation and configuration of large-scale magnetic fields in accretion disks remain a topic of active investigation. Seed fields may originate from local mechanisms such as the Biermann battery \citep{1982PASP...94..627K} or turbulent dynamos, or they may be advected from the external environment, including companion stars or the interstellar medium \citep{BisnovatyiKogan1974, 10.1093/mnras/stab3790}. However, the inward advection of magnetic flux is not a straightforward process, as MRI-driven turbulence also induces turbulent diffusivity that tends to transport magnetic flux outward, giving rise to the so-called “flux transport problem” \citep{10.1093/mnras/267.2.235, 2013SSRv..178..119B}. Whether an accretion disk can accumulate sufficient poloidal flux therefore depends on a delicate competition between inward advection and outward diffusion, both of which are strongly influenced by the disk’s thermodynamics and geometry. 
The efficiency of inward flux advection is regulated by key plasma properties, often parameterized by the magnetic Prandtl number $Pm=\eta/\nu$, and is generally more efficient in geometrically thick, ADAFs than in standard thin disks \citep{Narayan1994,Narayan2002,Ferreira2011}.
The net flux controls whether the disk remains in a weak-field configuration or enters a magnetically arrested state.
Two principal paradigms describe magnetized accretion flows: the Standard And Normal Evolution (SANE) state, where magnetic fields are present but not dominant, and the Magnetically Arrested Disk (MAD) state, in which magnetic flux saturates the inner disk and strongly suppresses accretion. 
This state is achieved when large-scale magnetic flux accumulates near the BH until the magnetic pressure balances the ram pressure of the inflowing matter \citep{Narayan2003,Igumenshchev2003}. 
Analytically, such flows are expected to develop a quasi-stationary structure with an inner supersonic zone, a transition zone, and an outer region where the field continues to grow \citep{BisnovatyiKogan1976}. It is important to emphasize, however, that the classical analyses by Bisnovatyi-Kogan \& Ruzmaikin were primarily restricted to nearly radial flows, in which magnetic field amplification is dominated by compressional effects. In those models, rotational shear plays no explicit dynamical role, and the resulting magnetic configurations cannot distinguish between Keplerian and sub-Keplerian angular momentum distributions. As a consequence, while these studies established the fundamental importance of flux freezing during gravitational infall, they do not address how rotational shear modifies magnetic field growth in realistic disk-like accretion flows, nor how such effects influence the approach to MAD states.
Throughout this work, the classical results of Bisnovatyi-Kogan \& Ruzmaikin (1974, Ref.~\citep{BisnovatyiKogan1974}) are recovered as a limiting case corresponding to vanishing rotational shear. This provides a consistency check for our formulation, while the inclusion of rotational effects leads to qualitatively new magnetic field evolution not captured in Ref.~\citep{BisnovatyiKogan1974}. Building on this perspective, the present work is specifically designed to fill this gap.

In the SANE regime, angular momentum transport is primarily mediated by MRI-driven turbulence, whereas in the MAD regime, the accumulation of magnetic pressure eventually counteracts the ram pressure of accretion, halting smooth inflow and rendering accretion highly intermittent and episodic \citep{2012MNRAS.423.3083M}. In the MAD state, the accumulation of large-scale poloidal magnetic flux near the black hole enables the launching of powerful relativistic jets via the Blandford–Znajek mechanism \citep{Igumenshchev2003, Blandford_Znajek_1977}. This process is extremely sensitive to the geometry of the magnetic field; for instance, the angle between field lines and the disk mid-plane must typically remain below $\lesssim60^\circ$ to ensure efficient jet launching from cold Keplerian disks \citep{Blandford1982,Cao1997}.
The critical threshold for MAD formation is defined by the dimensionless magnetic flux parameter $\phi$ threading the BH horizon, with GRMHD simulations identifying $\phi \gtrsim 50$ as the onset of magnetic arrest \citep{Tchekhovskoy_2011}. These simulations consistently reveal jet efficiencies exceeding 100\% of the accreted rest-mass energy, particularly for rapidly spinning black holes \citep{Tchekhovskoy_2011}.
Observational evidence provides strong support for these theoretical predictions: polarimetric imaging by the Event Horizon Telescope (EHT) reveals ordered magnetic field structures near the event horizons of M87 and Sgr A**, consistent with MAD-type configurations \citep{2021ApJ...910L..12E, EHT_SgrA_Polarization_2024}.
Furthermore, state-of-the-art GRMHD simulations have demonstrated complex, time-dependent MAD behavior, including quasi-periodic flux eruptions, turbulent mixing, and surface instabilities of magnetic flux tubes, all of which modulate energy dissipation and drive emission variability \citep{2018ApJ...855L..27P, Ripperda_2022}.
Recent radiation-GRMHD (GRRMHD) studies have extended this framework, examining how magnetically arrested states affect radiative efficiency, spectral formation, and thermal structure in luminous thin disks \citep{2020MNRAS.494.3656L}. While these numerical advances provide crucial insights into the global behavior of magnetized accretion, they also highlight inherent limitations: GRMHD simulations require extensive computational resources, and their complexity can obscure fundamental physical mechanisms. Moreover, key plasma microphysical processes—such as resistivity, reconnection, and particle acceleration—remain modeled only approximately, emphasizing the continuing importance of analytical and semi-analytical approaches that can clarify underlying physics, efficiently explore parameter space, and serve as benchmarks for numerical simulations.

In our previous work \citep{PaperI}, we introduced a semi-analytical framework targeting this very gap—within the Newtonian approximation. While these results clarified the role of angular momentum in the outer regions of accretion flows, they also highlighted the limitations of a purely Newtonian treatment. Crucially, in the relativistic regime close to the event horizon, spacetime curvature and relativistic azimuthal compression introduce shear-related effects that have no analogue in the classical treatments of Bisnovatyi-Kogan \& Ruzmaikin, making a general relativistic extension essential rather than incremental.

We analyzed how magnetic field amplification depends on angular momentum profiles in both Keplerian and sub-Keplerian flows and demonstrated that angular momentum structure crucially governs magnetic flux growth in outer disk regions. That work highlighted the importance of the balance between advection and diffusion in setting the flux accumulation efficiency. However, as one approaches the BH’s innermost regions, General Relativity (GR) becomes essential, and the Newtonian approximation of \citep{PaperI} does not capture relativistic corrections such as spacetime curvature, gravitational redshift, and crucial effects like relativistic azimuthal compression. This motivates extending the analysis into the fully relativistic domain.
To address this, in the present study we extend our semi-analytical formalism into the relativistic regime by modeling accretion flows in Schwarzschild spacetime. Specifically, under GR, we track the evolution of large-scale magnetic fields in both Keplerian and sub-Keplerian flow configurations. Our primary objectives are to determine: (i) which angular momentum profile is most efficient at accumulating the critical magnetic flux necessary to trigger the MAD state under strong gravity conditions, and (ii) what characteristic magnetic field strengths are achieved, and how they compare with findings from GRMHD simulations and observational constraints. By analyzing these questions, our model provides an intuitive and predictive framework that complements numerical studies and informs interpretations of high-resolution observations.

\section{Theoretical Framework}\label{sec:theoretical_framework}

\subsection{The Physical Model and Assumptions}\label{subsec:physical_model}

Accretion of magnetized plasma onto a black hole is a central problem in relativistic astrophysics, underlying energy release in quasars, X-ray binaries, and relativistic jets. Here, we adopt a semi-analytical approach to investigate how rotational shear affects magnetic field amplification in these flows. A self-consistent description must incorporate general relativistic effects alongside magnetohydrodynamic (MHD) interactions. In Schwarzschild spacetime, the gravitational field dictates infall trajectories, while angular momentum introduces deviations from pure free fall, generating rotationally supported structures and modifying radial profiles of velocity, density, and magnetic fields. While traditional analytic models often consider limiting cases such as purely radial infall or strictly Keplerian thin disks, observations and numerical simulations show that accretion flows frequently exhibit sub-Keplerian dynamics, particularly in hot, radiatively inefficient regimes like ADAFs and magnetically arrested disks (MADs).

To model this complex system, we adopt a set of foundational assumptions. We consider a stationary and axisymmetric flow, where the plasma dynamics are described by radial ($u^r$) and azimuthal ($u^\varphi$) velocity components, in addition to the temporal component ($u^t$). The inclusion of an azimuthal velocity component ($u^\varphi$), combined with the four-velocity normalization condition ($u^\mu u_\mu = 1$) in Schwarzschild spacetime, naturally leads to a radial velocity ($u^r$) that depends on the polar angle ($\theta$). This $\theta$-dependence reflects the combined effect of rotation and relativistic frame geometry on infall trajectories. The plasma is treated as a perfect conductor, which implies the ``frozen-in'' condition of ideal MHD. This assumption enforces vanishing resistivity and excludes reconnection or dissipative magnetic processes, ensuring that magnetic flux is conserved along each streamline. Although this assumption neglects reconnection, it isolates the fundamental amplification mechanisms from compressional and rotational effects. The resulting evolution thus isolates the amplification produced purely by compression and differential rotation. Mathematically, this is expressed by the vanishing of the electric field in the plasma's rest frame ($F^{\mu\nu}u_\nu=0$), meaning magnetic field lines are advected with the flow. Furthermore, the magnetic field four-vector is defined to be orthogonal to the four-velocity ($B^\mu u_\mu=0$), which reduces the independent field components to three, consistent with classical physics. We assume an initial large-scale poloidal magnetic field, containing only radial ($B^r$) and polar ($B^\theta$) components. Throughout this work we restrict ourselves to the dynamically weak-field regime, in which the magnetic energy density remains small compared to the gravitational binding energy of the flow. Consequently, the Lorentz-force back-reaction on the velocity field is neglected, and the plasma dynamics follow geodesic-like trajectories determined primarily by gravity and angular momentum. It is crucial to note that while the flow is stationary in the Eulerian frame (i.e., properties at a fixed spatial point are constant), the magnetic field is amplified in the Lagrangian frame as each fluid element is compressed and sheared during its infall. In this formulation the background flow remains stationary, but the magnetic field evolves non-stationarily along infalling trajectories. This separation is consistent with ideal-MHD treatments in strong gravity and ensures that temporal derivatives in the induction equation appear only through the Lagrangian evolution of individual fluid elements.

To provide a generalized framework for the flow's rotation, we express the azimuthal velocity as a fraction of the local Keplerian value:
\begin{equation}
\Omega = \xi \, \Omega_K, \quad 0 < \xi \le 1,
\label{eq:xi_definition}
\end{equation}
where $\xi=1$ represents a purely Keplerian disk and $\xi < 1$ corresponds to sub-Keplerian flows. Here, $\Omega_K = (r_g / r)^{3/2}/r_g$ denotes the \emph{local} Keplerian angular velocity of \emph{hypothetical} circular geodesics, even though no stable orbits exist for $r < 6 r_g$. This parameterization allows a continuous and physically intuitive variation of rotational support across all radii, providing a smooth transition between Keplerian and sub-Keplerian regimes. In the induction equations presented below, $\xi$ explicitly modulates the rotational shear term, controlling the relative amplification of radial and polar magnetic field components. This formalism also bridges the gap between idealized analytic models and GRMHD simulations by systematically exploring the effect of angular momentum on the magnetic field evolution. By employing the Schwarzschild metric, our model fully incorporates key relativistic corrections, including time dilation, gravitational redshift, and strong-field spacetime curvature, ensuring that the flow kinematics and magnetic evolution are treated consistently in the near-horizon region.

A fundamental distinction exists between the outer, quasi-Newtonian regions of the flow and the inner, strongly relativistic domain. As shown in our previous work \cite{PaperI}, the outer flow is typically subsonic with a long hydrodynamic timescale ($t_s \sim r^3$), resulting in a slowly evolving, nearly stationary zone. In contrast, the inner region analyzed here is characterized by rapid, supersonic infall ($v \sim r^{-1/2}$) and a much shorter hydrodynamic timescale ($t_s \sim r^{3/2}$). This accelerated inflow near the horizon is an inevitable feature of black hole accretion, even for rotationally supported plasma, and it drives a much faster evolution of the magnetic field structure. These distinct timescales motivate a semi-analytical approach that tracks the Lagrangian evolution of magnetic fields across both regimes. This physical picture provides a rigorous foundation for the governing equations derived in the following section.

\subsection{Governing Equations}\label{subsec:GoverningEquations}

To describe the dynamics of a magnetized accretion flow in the strong-gravity regime, we adopt the Schwarzschild metric for a non-rotating black hole:
\begin{equation}
    ds^{2} = \left(1-\frac{r_{g}}{r}\right)c^{2}dt^{2} - \left(1-\frac{r_{g}}{r}\right)^{-1}dr^{2} - r^{2}\left(d\theta^{2} + \sin^{2}\theta\, d\varphi^{2}\right),
    \label{eq:schwarzschild_metric}
\end{equation}
where $r_g = 2GM/c^2$ is the gravitational radius. In what follows, we adopt geometrized units where $G = M = c = 1$. The four-velocity of the plasma is given by $u^{\mu} = (u^t, u^r, 0, u^\varphi)$, consistent with the axisymmetric nature of the flow described in Section~\ref{subsec:physical_model}.

The evolution of the magnetic field is governed by the ideal MHD equations in curved spacetime. The covariant induction equation is derived from the source-free Maxwell equations:
\begin{equation}
    \partial_\gamma F_{\alpha \beta} + \partial_\alpha F_{\beta \gamma} + \partial_\beta F_{\gamma \alpha} = 0,
\end{equation}
where $F_{\alpha\beta} = u_\alpha b_\beta - u_\beta b_\alpha$ is the electromagnetic field tensor and $b^\mu$ is the magnetic field four-vector. For a stationary, axisymmetric inflow, the conservation of magnetic flux can be expressed as an integral form \citep{BisnovatyiKogan1974}:
\begin{equation}
    \sqrt{-g}\, u_t^{-1} B^r (u_r u^r + u_t u^t) = C_2,
    \label{eq:magFlux_r}
\end{equation}
\begin{equation}
    \sqrt{-g}\, u_r B^\theta = C_3,
    \label{eq:magFlux_theta}
\end{equation}
In the limit of vanishing rotational shear, these relations reduce exactly to the magnetic flux conservation laws derived by Bisnovatyi-Kogan \& Ruzmaikin \citep{BisnovatyiKogan1974}. The present formulation therefore recovers the classical solution as a limiting case, while retaining the full relativistic structure and allowing for differential rotation to modify the magnetic field evolution.
Where $g$ is the determinant of the metric tensor and $C_2, C_3$ are constants of integration. The trajectory of a fluid element is given by:
\begin{equation}
    ct - \int \frac{dr}{u^r/u^t} = C_1,
    \label{eq:magTime}
\end{equation}
where $C_1$ is another integration constant. This constant gives rise to a parameter, $x_0$, which possesses a fundamental physical interpretation. It is not merely a mathematical artifact of integration; rather, $x_0$ functions as a Lagrangian coordinate, serving as an invariant "label" for each spherically symmetric shell of plasma. A fluid shell that is located at the dimensionless radius $x_0 = r_g/r_0$ at the initial time ($t=0$) is uniquely identified by this parameter throughout its subsequent infall towards the event horizon. This property allows us to track the evolution of individual fluid elements. The introduction of $x_0$ thus provides a natural bridge between the Eulerian description of the flow and a Lagrangian viewpoint, which is particularly well suited for analyzing magnetic field amplification under strong radial compression near the event horizon.
These equations form the basis for computing the evolution of the poloidal magnetic field components for a given velocity profile. The complete solution procedure is detailed in Appendix~A of PaperI \citep{PaperI}.

\subsection{Model Setup and Initial Conditions}\label{subsec:initial_conditions}
Our analysis focuses on the strongly relativistic inner region ($r \lesssim 10\,r_g$), where plasma dynamics are dominated by the black hole’s gravity. We initialize the Lagrangian integration at a fiducial inner boundary of $r_0 = 3\,r_g$ (i.e., $x_0 = r_g/r_0 = 1/3$), placing the starting point well inside the innermost stable circular orbit (ISCO) at $r_{\rm ISCO} = 6\,r_g$. At this location, the flow is in the plunging, highly supersonic regime, yet retains significant rotational support parameterized by $\xi$ (with $\xi=1$ corresponding to the local Keplerian value and $\xi<1$ to sub-Keplerian rotation). The choice of $r_0 = 3\,r_g$ is motivated by the desire to isolate the relativistic plunging region, where radial compression and frame-drag-free gravitational effects dominate the dynamics, while simultaneously avoiding transitional effects associated with the ISCO. We have verified that the qualitative behavior of the magnetic field amplification is not sensitive to the precise value of the inner starting radius within this regime. This setup enables us to directly probe the magnetic field evolution under the combined influence of strong gravitational effects, rapid radial compression, and differential rotation — distinctly complementing the quasi-Newtonian outer-disk analysis presented in our previous work \cite{PaperI}.

\noindent The initial poloidal magnetic field at the inner boundary ($r_0 = 3\,r_g$) is adopted from the inwardly advected solution of our previous outer-disk study \cite{PaperI}. We set the physical components — measured in the local orthonormal frame of a zero-angular-momentum observer (ZAMO) — to $B_r(r_0) \simeq 30$ and $B_\theta(r_0) \simeq 0.1$, scaled consistently with the weak-field approximation. These values are intended to represent a predominantly radial magnetic configuration characteristic of advective inflows, without loss of generality for the qualitative amplification trends discussed below. This physically-motivated choice ensures continuity with the global accretion flow and allows us to isolate the additional relativistic amplification occurring in the near-horizon region. We emphasize that these values are representative and serve to illustrate the relative amplification and redistribution of the magnetic field components; the qualitative conclusions of the present study do not depend on the specific numerical normalization of the initial field.

\section{Magnetic Field Evolution in the Relativistic Inner Region}
\label{sec:RelativisticFlows}
In the innermost accretion region, within a few gravitational radii ($r \lesssim 10\,r_g$) of a non-rotating black hole, plasma dynamics are dominated by strong-field general relativistic effects \citep{Novikov1973, Shapiro_Teukolsky_1983, McKinney_2012}. The inflowing plasma experiences rapid radial acceleration, while azimuthal motion is constrained by the relativistic conservation of angular momentum. The curvature of spacetime significantly modifies both density profiles and the evolution of magnetic flux \citep{Gammie_2003, Tchekhovskoy_2011}, influencing jet launching conditions and the potential formation of magnetically arrested disks (MADs) \citep{Narayan_2012, McKinney_2014, EventHorizonTelescope_2019}.

\subsection{Relativistic Flow Kinematics and Trajectory}
\label{subsec:flow_kinematics}
The kinematics of the accreting plasma are determined by its four-velocity, $u^\mu$. For the axisymmetric flow defined in Section~\ref{subsec:physical_model} with an angular velocity parameterized by $\xi$, the components of the four-velocity in Schwarzschild coordinates are obtained from the normalization condition $u^\mu u_\mu = 1$, assuming the matter is at rest at infinity \citep{BisnovatyiKogan1974}:
\begin{equation}
    u^{t} = (1-x)^{-1},\quad u^{r} = - \left( x - \frac{x \xi^2 \sin^2 \theta}{r_g (1-x)} \right)^{1/2},\quad u^{\varphi} = \frac{\xi x^{3/2}}{r_g^{3/2} (1-x)},
\label{eq:four_velocity_components}
\end{equation}
For $\xi=0$, these expressions reduce to purely radial relativistic infall, reproducing the kinematics assumed in classical radial accretion models, while $\xi \neq 0$ introduces genuine rotational shear absent in those treatments.
Where $x = r_g/r$. This formulation allows us to investigate specific configurations, such as a sub-Keplerian flow ($\xi = 0.8$) and a purely Keplerian flow ($\xi = 1$).

To obtain an explicit time-dependent solution, we must first determine the trajectory of a fluid element, $x(t)$, by solving the integral given in Equation~\eqref{eq:magTime}. This yields the relation:
\begin{equation}
\begin{split}
    \frac{ct}{r_g} &+ a x^{-\frac{5}{2} } + b x^{-\frac{3}{2} } + c x^{-\frac{1}{2} }
    + d \ln \frac{1 - \sqrt{x}}{1 + \sqrt{x}} 
    + f \left( \ln \frac{1 - \sqrt{x}}{1 + \sqrt{x}} \right)^2 \\
    &= a x_0^{-\frac{5}{2} } + b x_0^{-\frac{3}{2} } + c x_0^{-\frac{1}{2} }
    + d \ln \frac{1 - \sqrt{x_0}}{1 + \sqrt{x_0}}
    + f \left( \ln \frac{1 - \sqrt{x_0}}{1 + \sqrt{x_0}} \right)^2,
\end{split}
\label{eq:time_evolution_integral}
\end{equation}
where the coefficients $a, b, c, d, f$ are determined by a numerical fit to the integral in Equation~\eqref{eq:magTime} (see Appendix~\ref{sec:appendixA}). For the asymptotic limit near the event horizon ($x \to 1$), this complex relation can be approximated to find an explicit expression for the Lagrangian coordinate $x_0$ in terms of the current position $x$ and time $t$:
\begin{equation}
    x_{0} \simeq 1-2\left(1-\sqrt{x} \right) e^{{ct}/{2fr_{g} }}.
\label{eq:x0_asymptotic}
\end{equation}
This asymptotic form explicitly captures the rapid convergence of fluid trajectories near the horizon, reflecting the exponential stretching of proper time relative to coordinate time in the strong-field regime.
This trajectory solution provides the kinematic background required to solve for the magnetic field evolution.

\subsection{Solution for the Magnetic Field Components}
\label{subsec:magnetic_field_solution}
By substituting the relativistic kinematics (Eq.~\ref{eq:four_velocity_components}) and the time-dependent trajectory (Eq.~\ref{eq:x0_asymptotic}) into the flux conservation laws (Eqs.~\ref{eq:magFlux_r} and~\ref{eq:magFlux_theta}), and applying the physically motivated initial conditions defined in Section~\ref{subsec:initial_conditions}, we obtain the explicit solution for the contravariant components of the magnetic field:
\begin{equation}
\begin{split}
    B^{\theta} = & -\frac{B_{0}}{r_{g}} \sin\theta  
    \left(1 - 2\left(1-\sqrt{x} \right) e^{{ct}/{2fr_{g} }} \right)^{-1/2}
    x^{{3}/{2}} \left(1 - x \right)^{1/2} \\
    & \times \left\{
    \frac{
        r_{g} \left[ 2\left(1-\sqrt{x} \right) e^{{ct}/{2fr_{g} }}  \right]
        - \xi^{2} \sin^{2}\theta
    }{
        r_{g}(1 - x) - \xi^{2} \sin^{2}\theta
    }
    \right\}^{1/2},\\
    B^r = & B_{0} \cos\theta 
    \left(1 -2\left(1-\sqrt{x} \right) e^{{ct}/{2fr_{g} }}  \right)^{-2}
    x^{2} (1 - x)^{2} \\
    & \times \left\{
    \frac{
        r_{g} \left(2\left(1-\sqrt{x} \right) e^{{ct}/{2fr_{g} }}  \right) ^{2}
        + \left(1 -2\left(1-\sqrt{x} \right) e^{{ct}/{2fr_{g} }}  \right) \xi^{2} \sin^{2}\theta
    }{
        \left(2\left(1-\sqrt{x} \right) e^{{ct}/{2fr_{g} }}  \right)^{3/2}
        \left( r_{g}(1 - x)^{2} + x\, \xi^{2} \sin^{2}\theta \right)
    } \right\}.
\end{split}
\label{eq:final_B_solution}
\end{equation}
In the limit $\xi \rightarrow 0$, these solutions reduce to the purely compressional amplification obtained in radial infall models, while finite $\xi$ introduces shear-driven terms that qualitatively modify the relative growth of $B^r$ and $B^\theta$.

Here, $B_0$ is a normalization constant related to the initial field strength. These equations describe the full time-dependent evolution of the poloidal magnetic field components as a function of radial position, time, and the angular momentum parameter $\xi$. 
It is essential to distinguish between the contravariant components of the magnetic field ($B^r, B^\theta$), which are coordinate-dependent quantities, and the physical components that would be measured by a local observer's magnetometer. In a curved spacetime, the coordinate basis vectors are not of unit length. The metric tensor components ($g_{\mu\nu}$) account for the local stretching and curvature of space, and they are used to project the field vector onto an observer's local orthonormal frame. Therefore, the physical components are obtained by scaling the contravariant components with the square root of the relevant metric components, such as in the relations $_rB = \sqrt{-g_{rr}}B^r$ and $_\theta B = \sqrt{-g_{\theta\theta}}B^\theta$.

% شکل 1
\begin{figure}[h]
\centering
\includegraphics[width=0.9\textwidth]{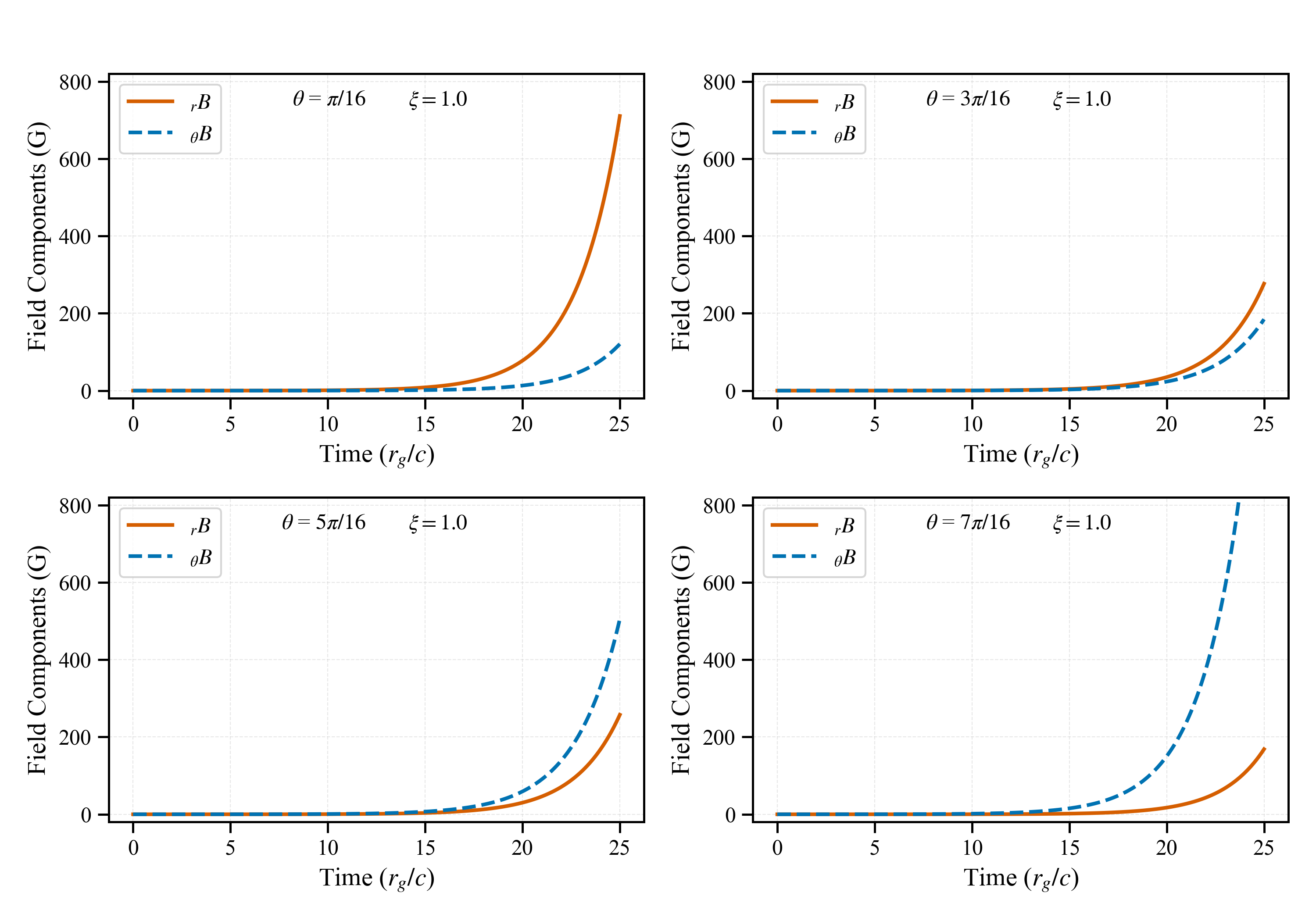}
\caption{Comparison of the time evolution of the physical components of the magnetic field for Keplerian flow ($\xi=1.0$) at different polar angles, measured at the fixed radial location $x=0.3$.}\label{fig:fig1}
\end{figure}

% شکل 2
\begin{figure}[h]
\centering
\includegraphics[width=0.9\textwidth]{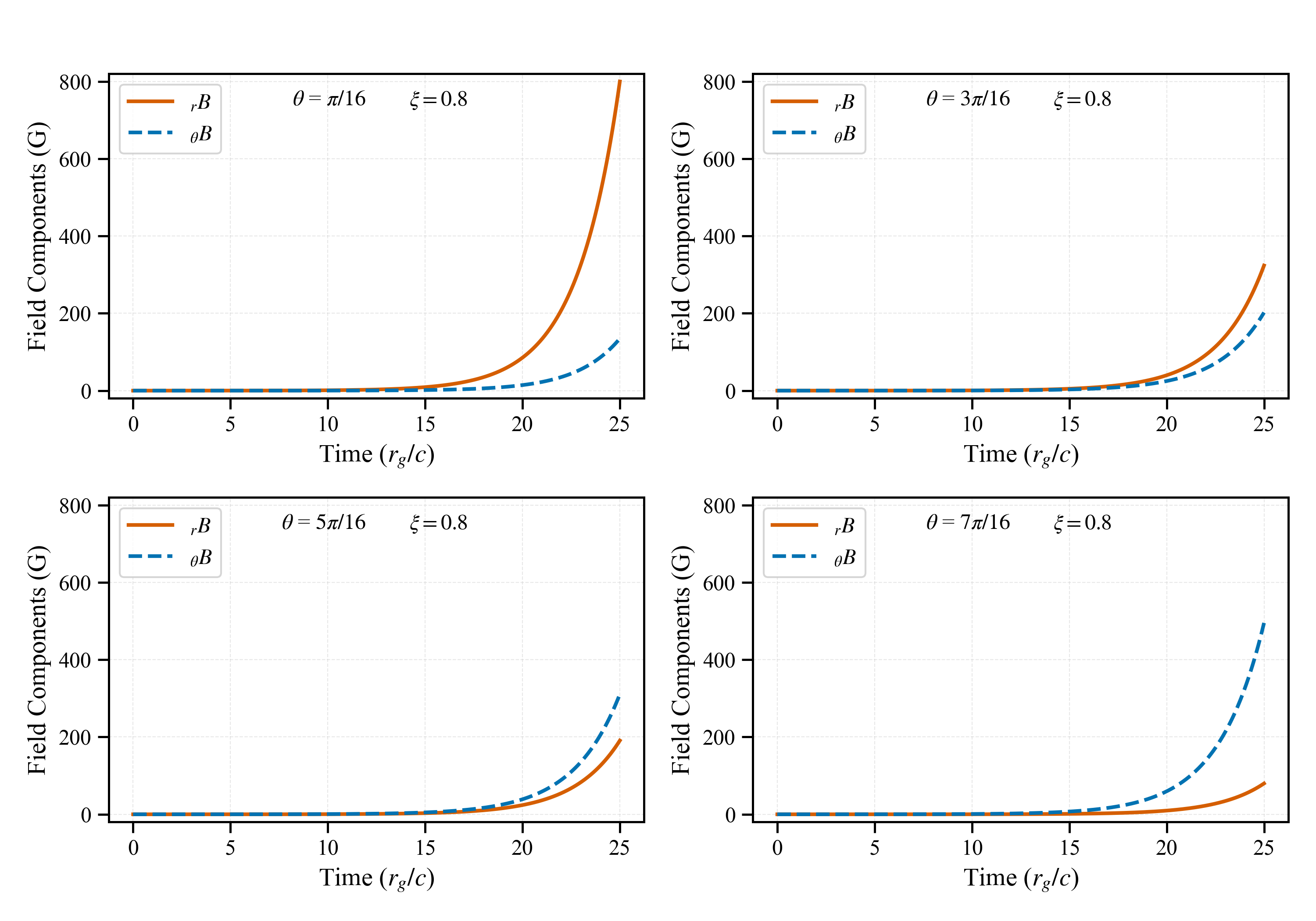}
\caption{Comparison of the time evolution of the physical components of the magnetic field for sub-Keplerian flow ($\xi=0.8$) at different polar angles, measured at the fixed radial location $x=0.3$.}\label{fig:fig2}
\end{figure}

% شکل 3
\begin{figure}[h]
\centering
\includegraphics[width=0.9\textwidth]{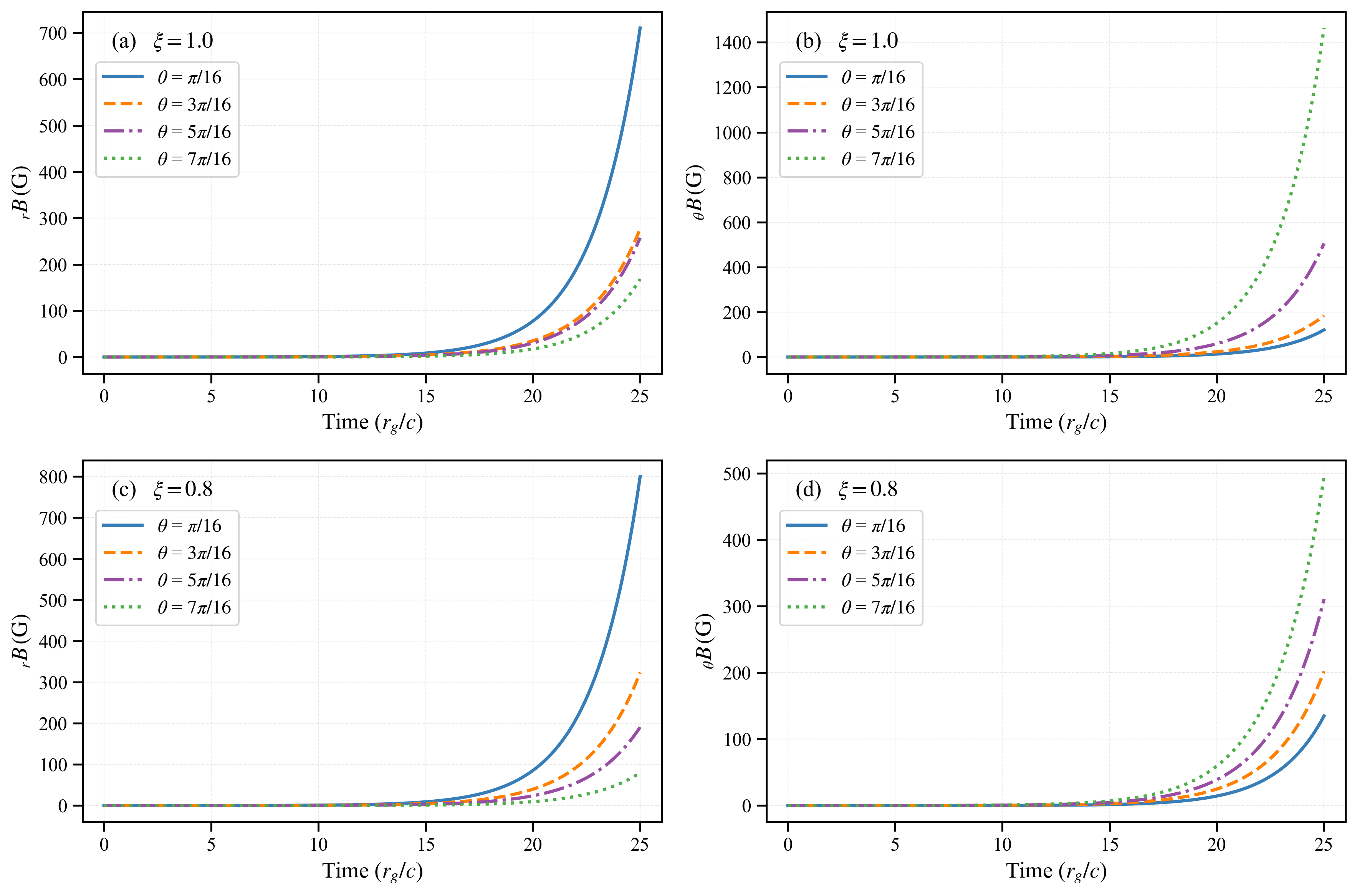}
\caption{Physical poloidal magnetic field components at the fixed strong-gravity location $x=0.3$ as a function of polar angle $\theta$, measured in the local orthonormal (ZAMO) frame.
(a) Radial component $_r B$ for Keplerian flow ($\xi=1.0$): slightly weaker than the sub-Keplerian case due to centrifugal barrier effects.
(b) Polar component $_\theta B$ for Keplerian flow ($\xi=1.0$): strongest near the equatorial plane ($\theta \approx 7\pi/16$), reflecting efficient amplification by relativistic rotational shear.
(c) Radial component $_r B$ for sub-Keplerian flow ($\xi=0.8$): significantly stronger across all angles, confirming radial compression as the dominant mechanism.
(d) Polar component $_\theta B$ for sub-Keplerian flow ($\xi=0.8$): substantially reduced compared to the Keplerian case owing to weaker differential rotation.
Panels (a,c) and (b,d) together illustrate the kinematic dichotomy: sub-Keplerian flows favor radial flux accumulation, whereas Keplerian flows maximize polar field winding.}
\label{fig:fig3}
\end{figure}

% شکل 4
\begin{figure}[h]
\centering
\includegraphics[width=0.9\textwidth]{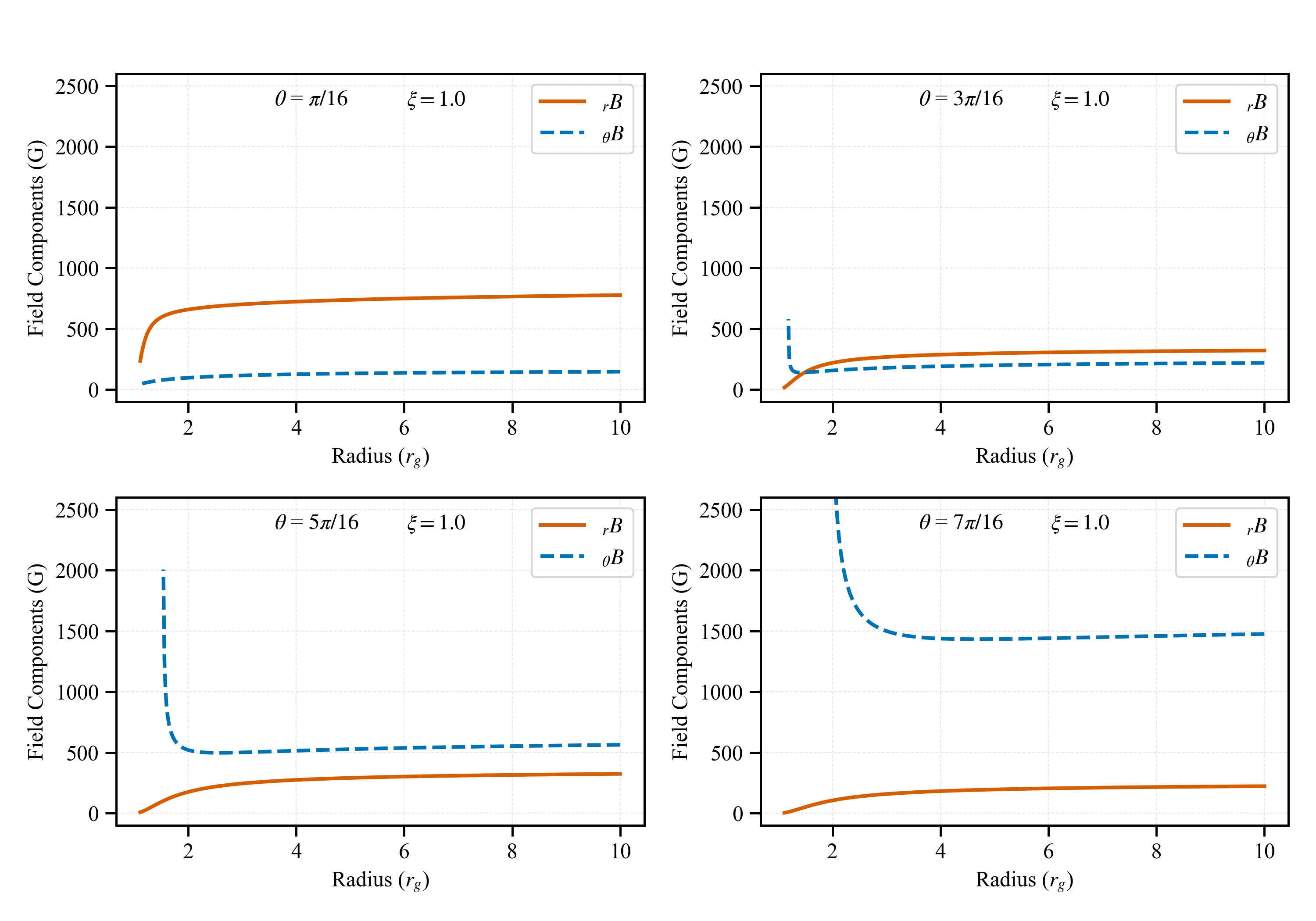}
\caption{Comparison of the variations in the radial ($_r B$) and polar ($_\theta B$) components of the Keplerian flow ($\xi=1.0$) magnetic field as a function of radius for different polar angles, measured at the fixed observer time $t=25\,r_g/c$.}\label{fig:fig4}
\end{figure}

% شکل 5
\begin{figure}[h]
\centering
\includegraphics[width=0.9\textwidth]{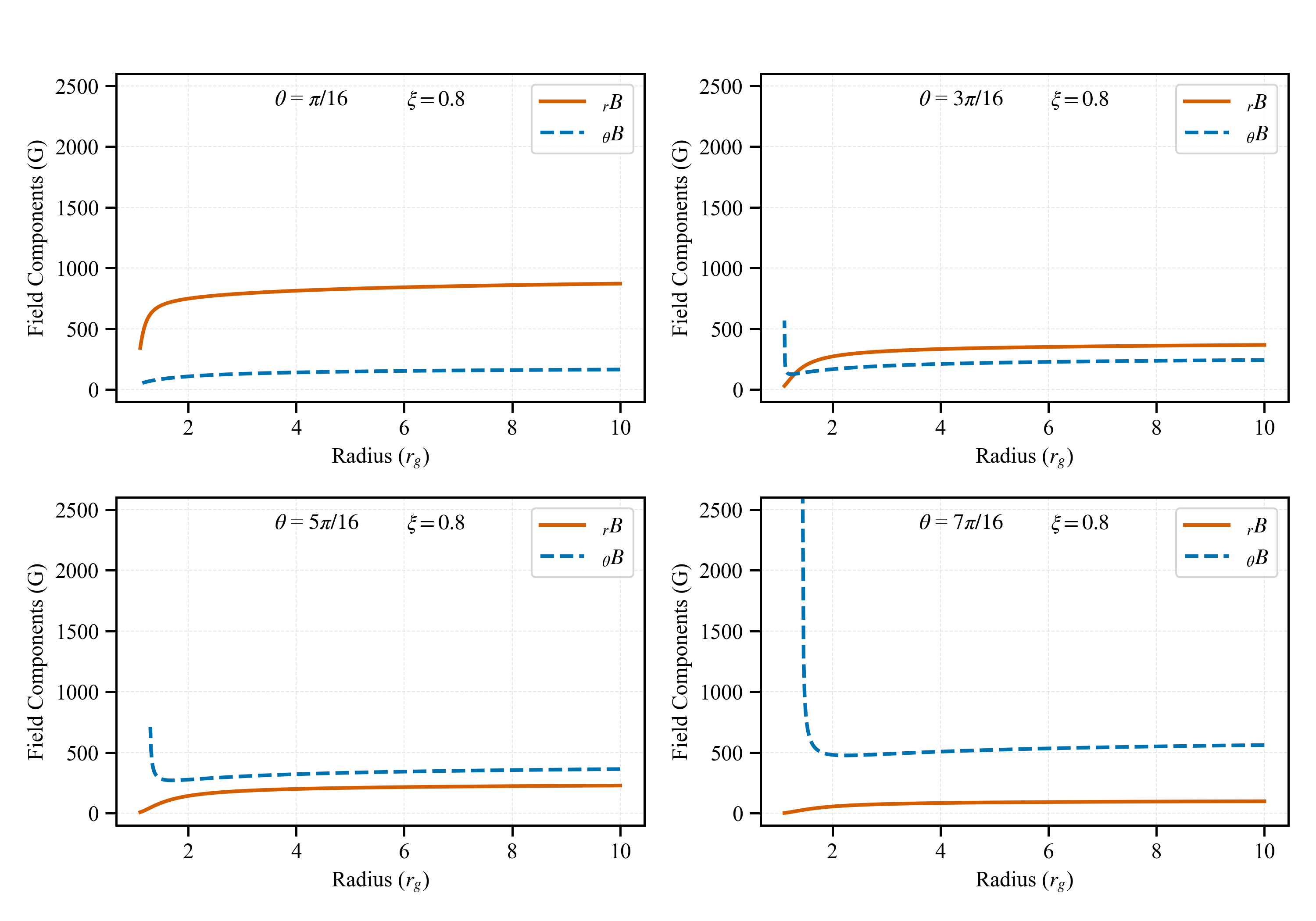}
\caption{Comparison of the variations in the radial ($_r B$) and polar ($_\theta B$) components of the sub-Keplerian flow ($\xi=0.8$) magnetic field as a function of radius for different polar angles, measured at the fixed observer time $t=25\,r_g/c$.}\label{fig:fig5}
\end{figure} 

% شکل 6
\begin{figure}[h]
\centering
\includegraphics[width=0.9\textwidth]{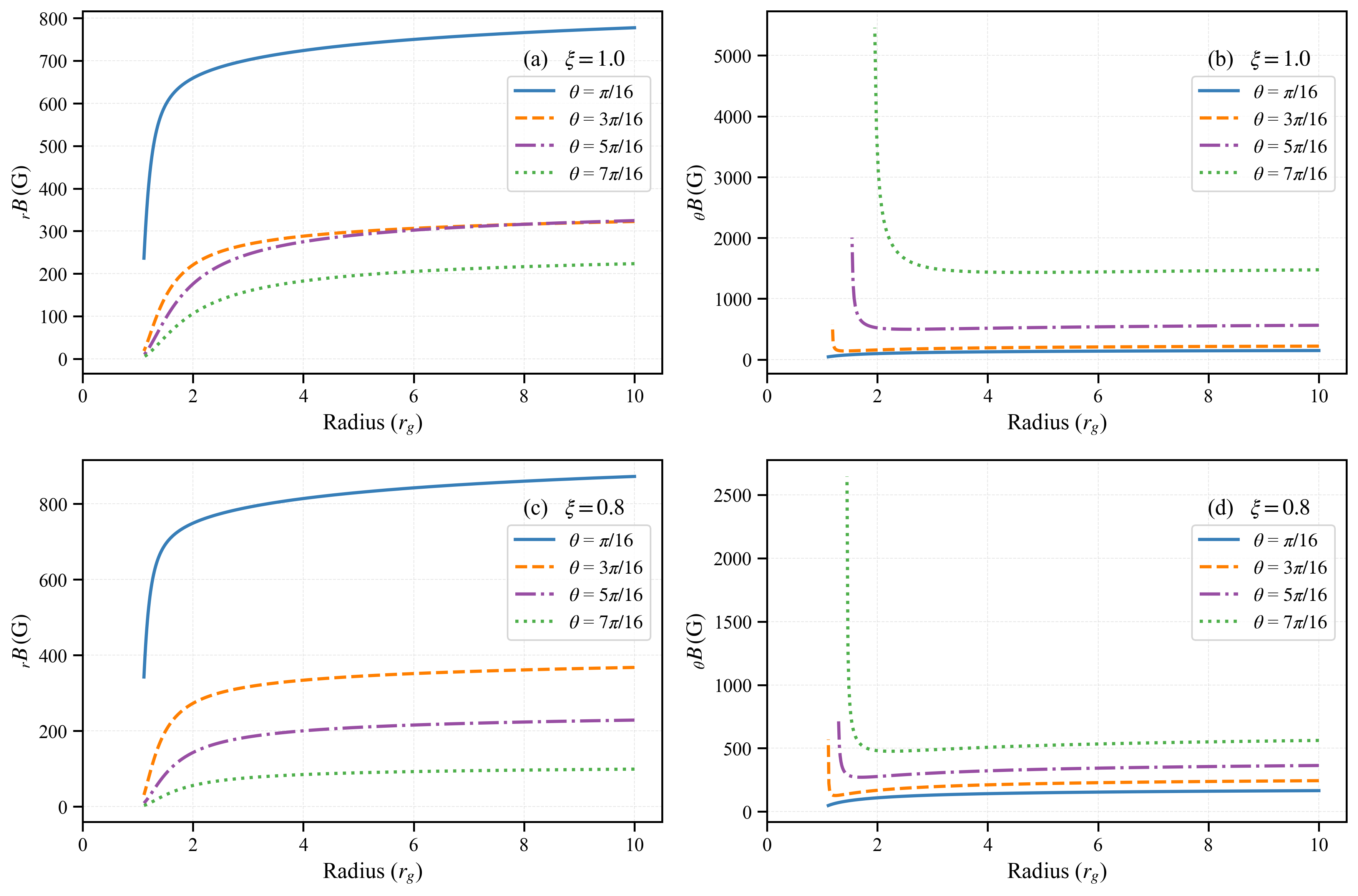}
\caption{Radial profiles of the physical poloidal magnetic field components at fixed observer time $t = 25\,r_g/c$, measured in the local ZAMO frame.
(a) Near the polar axis ($\theta = \pi/16$), Keplerian flow ($\xi=1.0$): $_r B$ is moderate, weaker than in sub-Keplerian flows due to centrifugal support, while $_\theta B$ (dashed) shows gradual growth.
(b) Near the equatorial plane ($\theta = 7\pi/16$), Keplerian flow ($\xi=1.0$):
$_\theta B$ strongly dominates in the innermost region ($r \le 3\,r_g$), reaching $\sim 5000\,G$, reflecting efficient amplification by relativistic rotational shear, while $_r B$ declines sharply.
(c) Near the polar axis ($\theta = \pi/16$), sub-Keplerian flow ($\xi=0.8$):
$_r B$ (solid) is stronger than in the Keplerian case, with $_\theta B$ remaining sub-dominant at large radii.
(d) Near the equatorial plane ($\theta = 7\pi/16$), sub-Keplerian flow ($\xi=0.8$):
Both components are reduced compared to the Keplerian case, but $_\theta B$ strongly exceeds $_r B$ in the innermost region ($r \le 3\,r_g$), confirming the transition to a non-radial structure.
All panels confirm the consistent transition to a highly non-radial, shear-dominated field configuration near the horizon, where $_\theta B \gg _r B$ in the equatorial region for both flows, which is essential for angular momentum transport.}
\label{fig:fig6}
\end{figure}

\section{Results: Magnetic Field Amplification and Structure in the Strong-Gravity Regime}
\label{sec:Results}
In this section, we present the results of our semi-analytical model, detailing the evolution and spatial structure of the physical magnetic field components ($_r B,_\theta B$) in the strong-gravity regime surrounding a non-rotating Schwarzschild black hole. These results are obtained within the relativistic extension of the Bisnovatyi-Kogan \& Ruzmaikin framework, where magnetic field amplification is driven by the combined effects of radial compression and differential rotation in a converging flow. In the strong-gravity regime considered here, these mechanisms operate on the local dynamical timescale, leading to rapid and highly efficient field growth.
A quantitative comparison with the Newtonian regime (reported in \citep{PaperI}) is provided to highlight the impact of General Relativity. All quantitative comparisons are derived from peak values at the final time step in the relativistic regime ($t=25\,r_g/c$) versus the saturation values in the Newtonian regime ($t=25000\,r_g/c$).

\subsection{Temporal Evolution and Growth Dynamics}
\noindent Figure~\ref{fig:fig1} illustrates the temporal evolution of the magnetic field components for the Keplerian flow ($\xi=1.0$) at a fixed radial position deep within the strong-gravity regime ($x=r_g/r=0.3$). The magnetic field components exhibit an initial quiescent phase, followed by a rapid, non-linear (approximately exponential) growth phase starting around $t \approx 15\,r_g/c$. This acceleration underscores the fact that field amplification in the relativistic regime is governed by the short \textit{dynamical timescale} near the horizon, a profound departure from the slower accretion timescale characterizing Newtonian flows. This behavior is a direct relativistic analogue of the compression-driven amplification identified by Bisnovatyi-Kogan \& Ruzmaikin, now strongly enhanced by the gravitationally induced convergence of the flow near the horizon. The subsequent quasi-periodic oscillations are indicative of magnetic flux accumulation and intermittent release via reconnection processes driven by strong differential rotation near the Event Horizon.
As depicted in Figure~\ref{fig:fig1}, the polar component ($_\theta B$) displays the most significant growth, peaking near the equatorial plane ($\theta = 7\pi/16$) at a substantial magnitude of $_\theta B \approx 1500\, \mathrm{G}$ at $t=25\,r_g/c$. This strong amplification of the $_\theta B$ component near the disk plane confirms the pre-eminence of \textit{relativistic rotational shear} as the key amplification mechanism for Keplerian flows, thus driving the magnetic configuration towards a highly non-radial and strongly sheared state. The impact of General Relativity is quantified by comparing these values to the corresponding Newtonian results \cite{PaperI}: $_\theta B$ at $\theta=7\pi/16$ is amplified by a factor of approximately 22000, while the radial component ($_r B$) at $\theta=\pi/16$ is amplified by a factor of approximately 21.5. This dramatic increase in magnitude, coupled with the drastic reduction in the growth timescale from $t \sim 25000\,r_g/c$ (as observed in \cite{PaperI}) to $t \sim 25\,r_g/c$, is a defining hallmark effect of the strong-gravity regime.

Figure~\ref{fig:fig2} presents the temporal evolution of the magnetic field components for the sub-Keplerian flow ($\xi=0.8$) at $x=r_g/r=0.3$. Due to the reduced angular momentum compared to the Keplerian case, the efficiency of rotational shearing decreases, leading to a weaker growth of $_\theta B$. Conversely, the resulting enhanced \textit{radial advection} of plasma more effectively drags magnetic flux inward. This mechanism manifests as sharper, more intense magnetic field \textit{spikes} near the horizon, related to episodic flux pile-up and subsequent reconnection bursts characteristic of magnetically active accretion flows. The angular dependence reflects this dynamical dominance: the radial component ($_r B$) peaks near the axis ($\theta = \pi/16$) at $_r B \approx 800\, \mathrm{G}$, a value $\approx 12\%$ higher than the Keplerian maximum. Conversely, the peak $_\theta B$ ($\approx 490\, \mathrm{G}$ at $\theta=7\pi/16$) is approximately one-third of the Keplerian value. This strong enhancement of $_r B$ confirms that \textit{radial compression} is the dominant $_r B$ amplification mechanism in sub-Keplerian flows. This trend closely follows the classical expectation for low-angular-momentum accretion, where magnetic flux freezing combined with rapid infall naturally favors poloidal field amplification. This favors a \textit{poloidal-dominated structure} near the black hole axis relative to the Keplerian flow, which is essential for launching relativistic outflows. The superiority of the strong-gravity regime in flux advection is again quantified through the comparison with Newtonian data \cite{PaperI}: $_r B$ (at $\theta=\pi/16$) is amplified by a factor of approximately 23.5, and $_\theta B$ (at $\theta=7\pi/16$) is amplified by a factor of approximately 7100. The greater $_r B$ amplification compared to the Keplerian case highlights the superior efficiency of flux advection enforced by the relativistic environment on less rotationally supported flows.

Figure~\ref{fig:fig3} quantitatively confirms the primary findings regarding the role of angular momentum $\xi$ in dictating the poloidal magnetic field component structure at the fixed strong-gravity location $x=r_g/r=0.3$. The multi-panel figure compares the field components as a function of polar angle $\theta$ for Keplerian ($\xi=1.0$) and sub-Keplerian ($\xi=0.8$) flows. Panel (a) illustrates the radial component $_r B$ for the Keplerian flow ($\xi=1.0$), which is generally weaker compared to the sub-Keplerian case, confirming that the centrifugal support effectively suppresses the direct radial advection of magnetic flux. Conversely, Panel (b) shows that the polar component $_\theta B$ for the Keplerian flow is significantly stronger, peaking near the equatorial plane. This high amplitude (nearly three times the peak of the sub-Keplerian case) confirms that rotational shearing is maximized in rotationally supported flows, efficiently winding the poloidal field lines. Furthermore, Panel (c) presents the radial component $_r B$ for the sub-Keplerian flow ($\xi=0.8$). This component is consistently the strongest across all polar angles, highlighting that faster radial infall and compression is the principal driver for the radial field component when rotational support is reduced. Finally, Panel (d) displays the polar component $_\theta B$ for the sub-Keplerian flow, which exhibits a substantially reduced magnitude compared to the Keplerian flow due to the weaker differential rotation. Together, these panels clearly illustrate the kinematic dichotomy: sub-Keplerian flows favor radial flux accumulation and enhance $_r B$, whereas Keplerian flows maximize the polar field winding (shear) and substantially enhance $_\theta B$.

\subsection{Analysis of Radial Profiles and Spatial Structure}

\noindent Figure~\ref{fig:fig4} presents the radial dependence of the magnetic field components for the Keplerian accretion flow ($\xi=1.0$) at $t=25\,r_g/c$. The data reveal a critical magnetic field reconfiguration in the innermost region ($r \le 4\,r_g$). As the fluid approaches the Event Horizon, the radial component, $_r B$, systematically decreases across all angles ($\theta$). For instance, near the pole ($\theta=\pi/16$), $_r B$ drops significantly from $870\, \mathrm{G}$ at $r=10\,r_g$ to $220\, \mathrm{G}$ at $r=2\,r_g$. This decline suggests a deflection of the poloidal flux lines away from the radial direction, primarily due to the intense relativistic gravitational stretching and confinement near the horizon. Conversely, the polar component, $_\theta B$, increases asymptotically, reaching its maximum amplification near the horizon, especially towards the equator ($\theta=7\pi/16$), where it increases from $1500\, \mathrm{G}$ to $6000\, \mathrm{G}$ in the range $10-2\,r_g$. This inverse radial correlation between $_r B$ depletion and $_\theta B$ amplification confirms a strong transformation to a non-radial, shear-dominated configuration in the inner disk. The dramatic asymptotic growth of $_\theta B$ is the direct signature of efficient \textit{magnetic flux pile-up} and overwhelming rotational shear, consistent with a quasi-steady structure where the differential rotation effectively wraps the field lines just outside the Event Horizon, leading to \textit{magnetic saturation}.

Following this, Figure~\ref{fig:fig5} shows the radial profiles for the sub-Keplerian flow ($\xi=0.8$). The structural trend observed is similar but accentuated by the enhanced radial infall velocity characteristic of this flow type. While $_r B$ also decreases sharply towards $r=2\,r_g$, $_\theta B$ exhibits steep asymptotic growth towards the horizon (e.g., $_\theta B \to 2600\, \mathrm{G}$ at $\theta=7\pi/16$). The \textit{steeper radial gradient} and highly compact magnetic structure compared to the Keplerian case reflect the superior efficiency of flux advection and radial compression in the inner region, where the rapid infall timescale dominates over the rotational shear timescale. Crucially, the simultaneous $_r B$ suppression and $_\theta B$ growth confirm that the strong gravitational compression and fluid convergence near the horizon are sufficient to geometrically focus and intensify the magnetic flux, leading to a compact, dynamically dominant structure believed to power transient relativistic jets and episodic ejections. 

Figure~\ref{fig:fig6} provides a direct comparison of the radial profiles of the poloidal magnetic field components, $_r B$ and $_\theta B$, plotted simultaneously for a specific polar angle in each panel. The figure covers both Keplerian ($\xi=1.0$, Panels a and b) and sub-Keplerian ($\xi=0.8$, Panels c and d) flows. Panel (a) shows the radial component $_r B$ for the Keplerian flow ($\xi=1.0$) across various angles; this component is weaker than the polar component, especially in the innermost region, due to centrifugal support. Panel (b) displays the polar component $_\theta B$ for the Keplerian flow, which exhibits strong dominance and efficient amplification in the innermost region ($r \le 3\,r_g$), particularly near the equatorial plane, confirming the effect of relativistic rotational shear. Furthermore, Panel (c) illustrates the radial component $_r B$ for the sub-Keplerian flow ($\xi=0.8$). While generally stronger than the Keplerian $_r B$ at large radii, it still drops sharply and remains sub-dominant compared to the sub-Keplerian $_\theta B$ near the horizon. Finally, Panel (d) presents the polar component $_\theta B$ for the sub-Keplerian flow; similar to the Keplerian case, this component strongly dominates $_r B$ in the strong-gravity regime, though the overall magnitude is reduced due to weaker differential rotation. The plots visually confirm the key structural finding: for both flows, the profiles consistently show that the polar component ($_\theta B$) strongly dominates the radial component ($_r B$) in the innermost region ($r \le 3\,r_g$), with $_r B$ dropping sharply. This spatial configuration demonstrates that close to the black hole, the magnetic field lines are heavily wrapped and confined by the fluid flow, resulting in a highly non-radial poloidal structure essential for supporting angular momentum transport via magnetic shear in the inner accretion flow.

\subsection{Impact of Angular Momentum on Amplification Hierarchy}

\begin{figure}[ht!]
    %\centering 
    \begin{subfigure}[b]{0.49\textwidth} 
       % \centering
        \includegraphics[width=\textwidth]{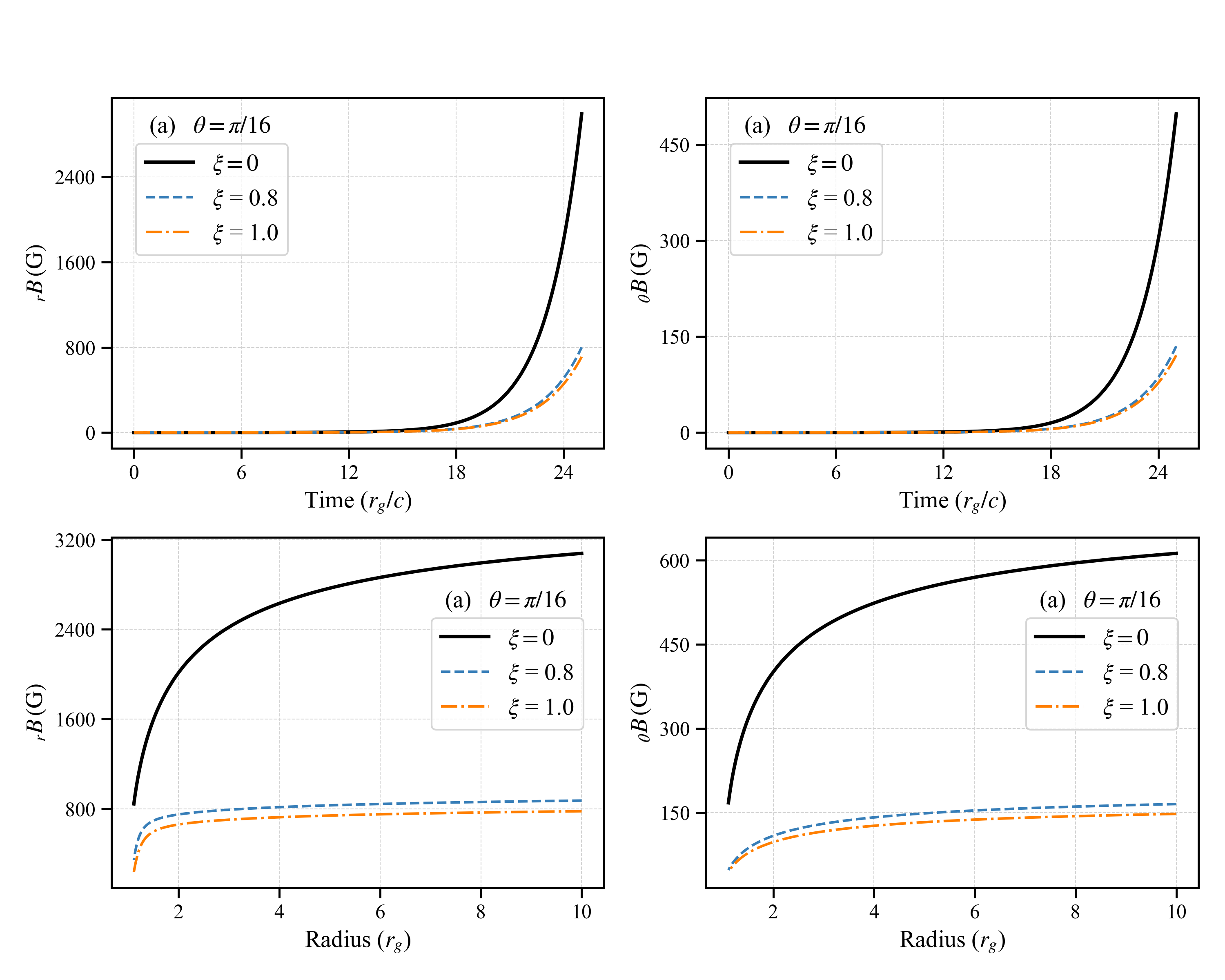}
        \phantomcaption
    \label{fig:fig7a}
    \end{subfigure}
    \hfill % Add horizontal space between subfigures
    \begin{subfigure}[b]{0.49\textwidth}
        %\centering
         \includegraphics[width=\textwidth]{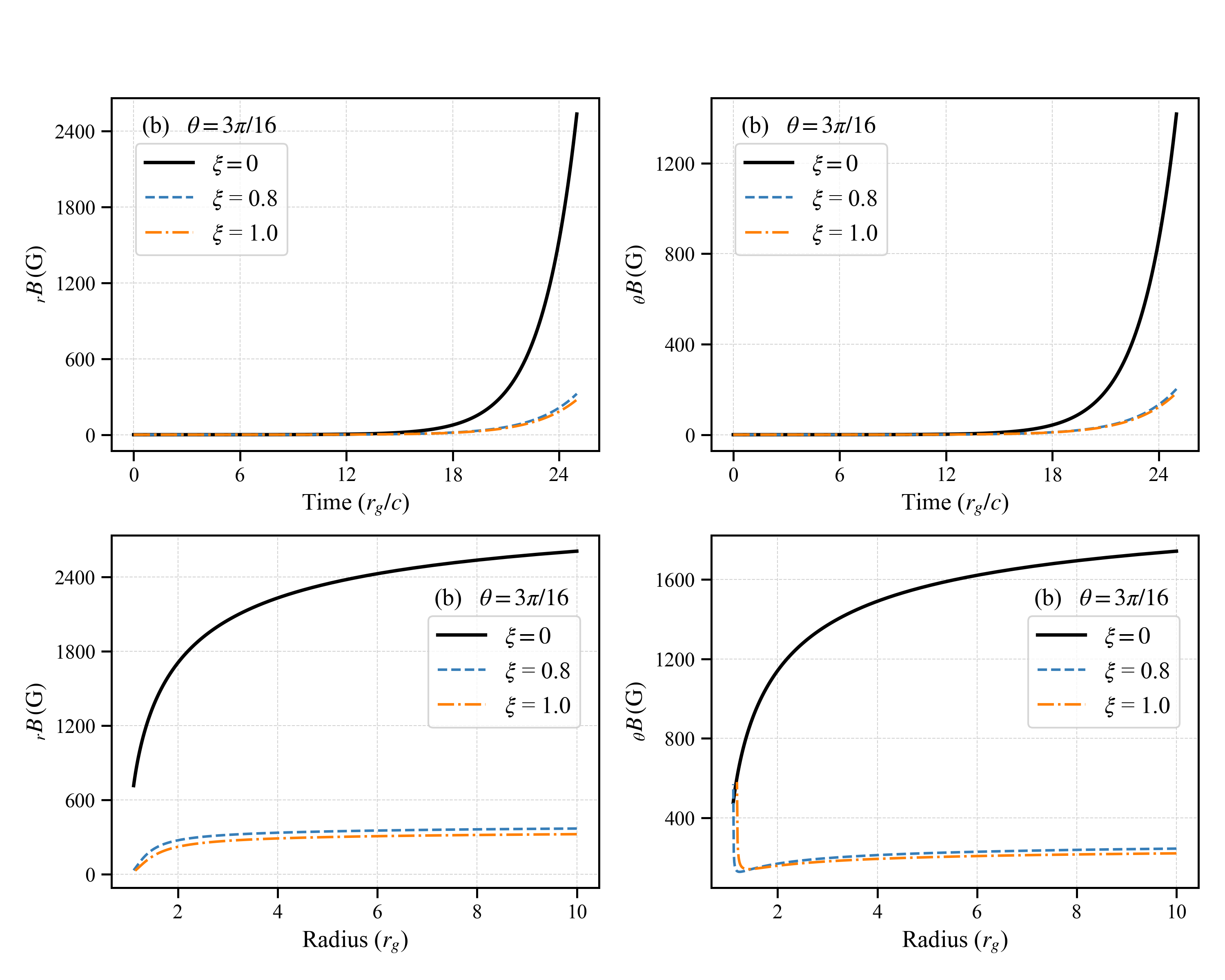}
        \phantomcaption 
    \label{fig:fig7b}
    \end{subfigure}

    \vspace{0.01cm} % Optional: add some vertical space between rows of subfigures

    \begin{subfigure}[b]{0.49\textwidth}
        \centering
         \includegraphics[width=\textwidth]{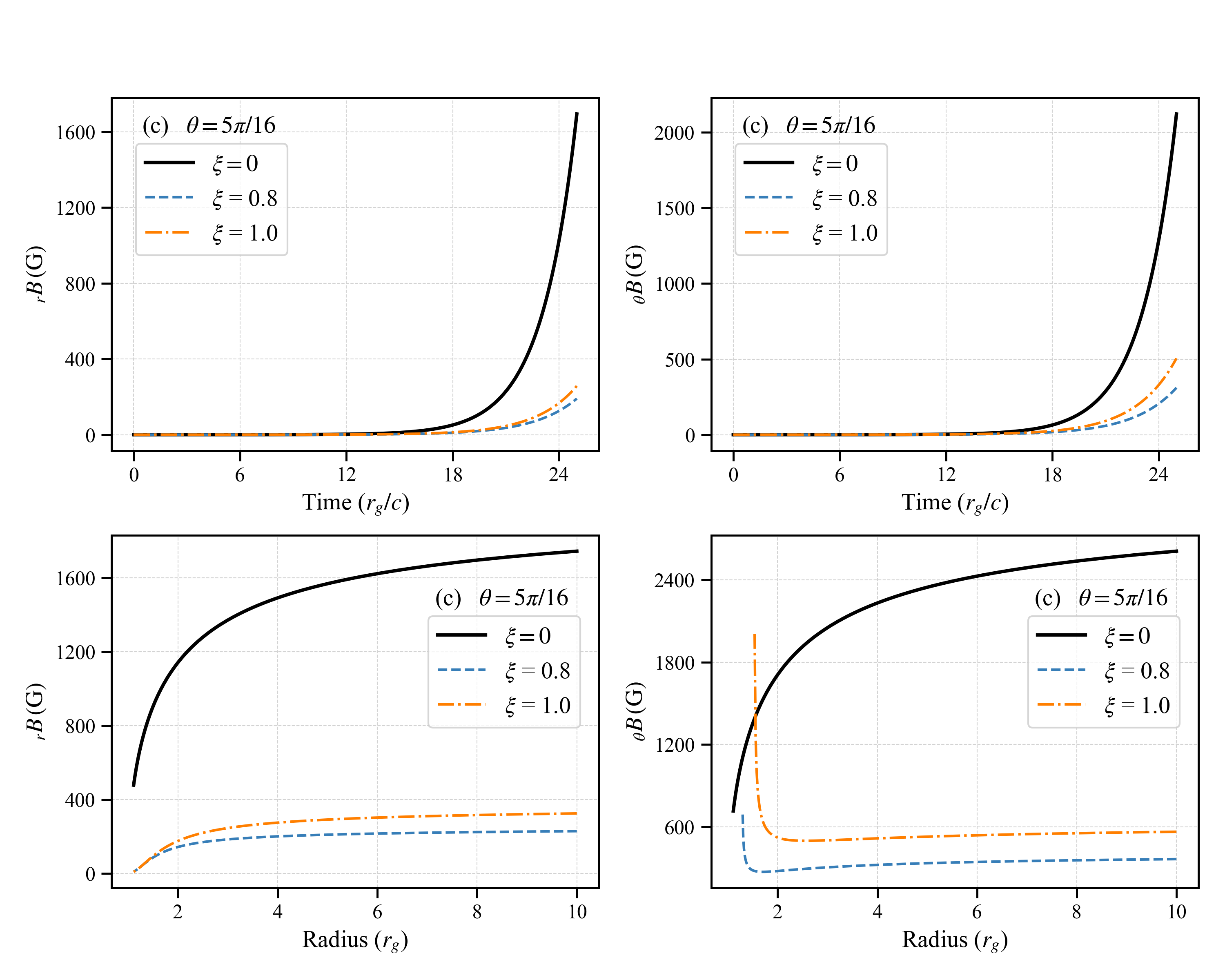}
        \phantomcaption 
    \label{fig:fig7c}
    \end{subfigure}
    \hfill
    \begin{subfigure}[b]{0.49\textwidth}
        \centering
         \includegraphics[width=\textwidth]{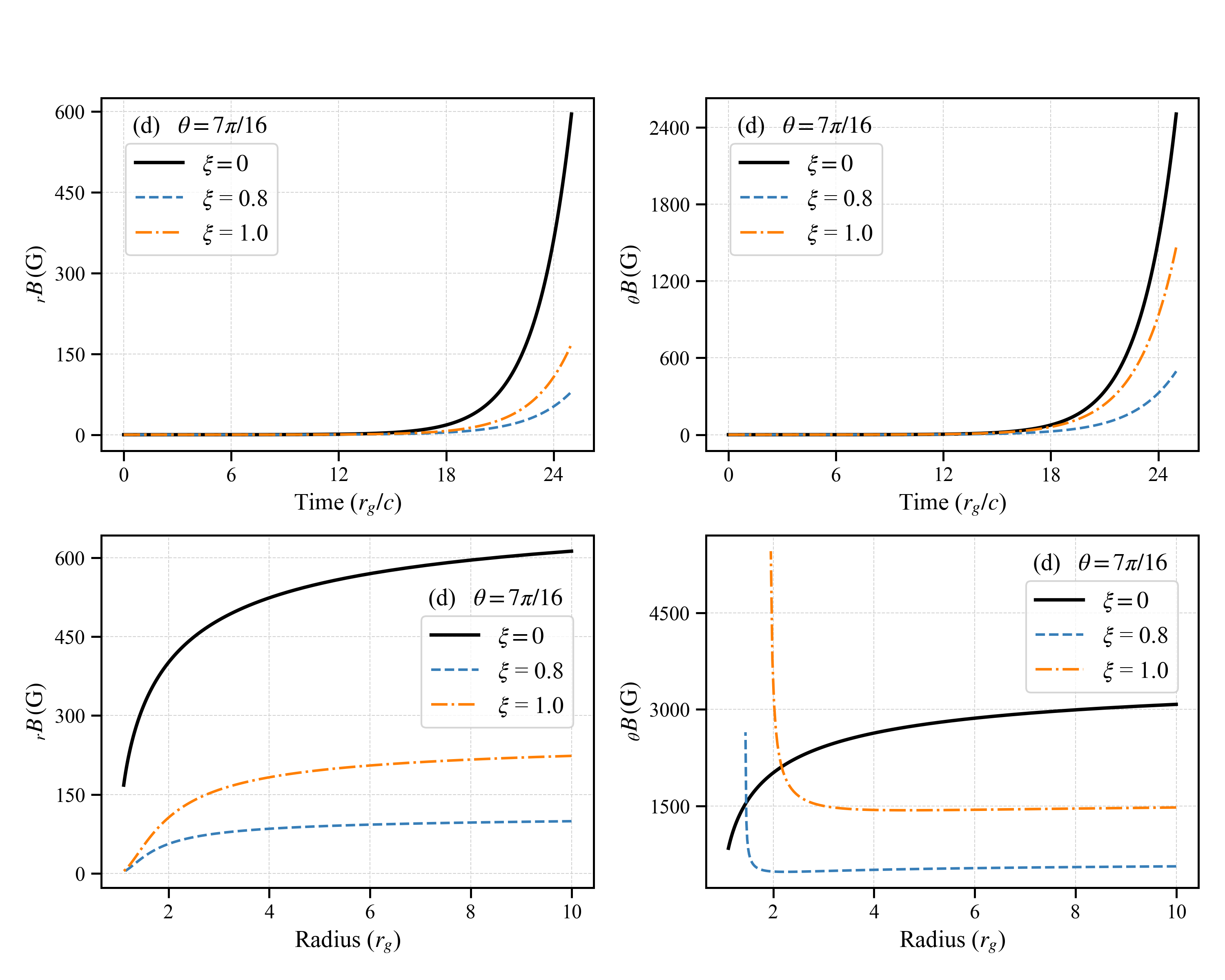}
        \phantomcaption 
    \label{fig:fig7d}
    \end{subfigure}
\caption{Comprehensive comparison of magnetic field amplification across three flow types (pure radial infall $\xi=0$, sub-Keplerian $\xi=0.8$, and Keplerian $\xi=1.0$) as a function of polar angle $\theta$. The four plots in each panel show the time and radial profiles of both components, measured at the fixed inner-boundary location $x=0.3$ and observer time $t=25\,r_g/c$. 
(a) Profiles near the polar axis ($\theta = \pi/16$): Highlights the dominance of radial compression, resulting in the hierarchy $\xi=0 \gg \xi=0.8 > \xi=1.0$ for both $_r B$ and $_\theta B$ components. 
(b) Profiles near $\theta = 3\pi/16$: Shows the increasing influence of rotational shear, where Keplerian flow begins to show a relative advantage over sub-Keplerian flow for $_\theta B$. 
(c) Profiles near $\theta = 5\pi/16$: Confirms the strong $_\theta B$ amplification hierarchy $\xi=0 \gg \xi=1.0 > \xi=0.8$ due to superior differential rotation in the Keplerian case. 
(d) Profiles near the equatorial plane ($\theta = 7\pi/16$): Provides the clearest signature of shear dominance, where the Keplerian flow significantly maximizes $_\theta B$ relative to the sub-Keplerian case. 
Together, the panels illustrate the transition from compression-dominated dynamics near the pole to shear-dominated dynamics near the equator.}
\label{fig:fig7}

\end{figure}

\noindent Figure~\ref{fig:fig7} provides a comprehensive comparison of the magnetic field amplification across three distinct flow types: purely non-rotating radial infall ($\xi=0$), sub-Keplerian flow ($\xi=0.8$), and Keplerian flow ($\xi=1.0$), with each panel (a-d) presenting the full time and radial profiles of $_r B$ and $_\theta B$ for a fixed polar angle $\theta$. This analysis clearly establishes the fundamental, monotonic inverse relationship between angular momentum ($\xi$) and magnetic field amplification. The non-rotating flow ($\xi=0$) is overwhelmingly the most efficient amplifier of both components, confirming that direct radial compression is the primary mechanism for generating the poloidal flux, a process severely suppressed by rotational support.
\noindent The four panels illustrate how the amplification hierarchy evolves from the pole to the equator:
Panel (a) ($\theta=\pi/16$, near the polar axis) shows the strongest signature of radial compression dominance: the hierarchy for the radial component, $_r B$, is clearly $\xi=0 \gg \xi=0.8 > \xi=1.0$. The polar component $_\theta B$ also follows this hierarchy, as rotational shear is minimal near the axis.
Panel (b) ($\theta=3\pi/16$) and Panel (c) ($\theta=5\pi/16$) reveal the growing influence of rotational shearing as the angle moves towards the equator. While $\xi=0$ remains dominant for both components, the Keplerian flow ($\xi=1.0$) begins to show a more pronounced advantage over the sub-Keplerian flow ($\xi=0.8$) in amplifying the polar component $_\theta B$, resulting in the hierarchy $\xi=0 \gg \xi=1.0 > \xi=0.8$ for $_\theta B$.
Finally, Panel (d) ($\theta=7\pi/16$, near the equatorial plane) provides the clearest confirmation of the dual role of rotation. Here, the $_\theta B$ profile for the Keplerian flow ($\xi=1.0$) shows the greatest relative advantage over the sub-Keplerian case, demonstrating the superior differential rotational shear effect where angular momentum density is highest. The radial profiles shown in these panels confirm the structural implication that the non-rotating ($\xi=0$) case maintains a strong poloidal-dominated structure, while the rotating flows transition to a highly non-radial structure where $_\theta B$ dominates $_r B$ near the horizon due to efficient field line wrapping.

% شکل  7

\section{Discussion: Magnetic Field Configuration, Angular Momentum, and Relativistic Dynamics}\label{sec:discussion}
Our relativistic analysis reveals profound enhancement of magnetic field amplification compared to the quasi-Newtonian regime explored in our previous work \cite{PaperI}. Key quantitative differences include 1.5–3 times stronger poloidal fields near the horizon, an accelerated stationary zone expansion ($r_{\rm stat}\propto t^{2/3}$ versus $t^{1/3}$ in the Newtonian domain \cite{PaperI}), as inferred from the present semi-analytical model and a clearer dichotomy in angular momentum effects. The sub-Keplerian flow ($\xi<1$) optimizes radial compression, while the Keplerian flow ($\xi=1.0$) maximizes shear amplification. These trends arise from the transition to supersonic infall and the causal boundary of the event horizon, which trap and intensify magnetic flux far more efficiently than in Newtonian gravity.

\subsection{Physical Interpretation of Flow Kinematics and Field Amplification}\label{subsec:discussion_physics}

Our results reveal three robust trends for the evolution of the poloidal magnetic field components in the relativistic inner flow (Figure~\ref{fig:fig7}): (i) a non-rotating inflow ($\xi=0$) amplifies both $_r B$ and $_\theta B$ most efficiently; (ii) a sub-Keplerian flow ($\xi=0.8$) more strongly amplifies the radial component $_r B$ than a Keplerian flow ($\xi=1.0$); and (iii) the Keplerian flow more strongly amplifies the polar component $_\theta B$. This clear hierarchy provides a unified physical explanation rooted in the competition between fundamental relativistic timescales.

\paragraph{Compression-dominated amplification at $\xi=0$.}
In the absence of rotation, centrifugal support vanishes, and gravity drives rapid radial convergence of streamlines. Under ideal MHD (frozen-in) conditions, flux conservation forces the poloidal field to intensify as the flow collapses. Pure radial compression steepens $_r B$ (flux tube cross-sections shrink with radius), while geometric focusing in curved spacetime tilts and concentrates field lines, contributing to $_\theta B$ growth. Physically, as streamlines converge towards the black hole's event horizon---a geometric focal point and a causal boundary---the frozen-in field lines are inevitably squeezed together both radially and latitudinally. With no shear ($d\Omega/d\ln r=0$), the amplification is set almost entirely by the short compression timescale ($t_{\rm comp}\sim r/|v_r|$), leading to the steep, overwhelming growth seen in Figure~\ref{fig:fig7}. The resulting structure, where $_r B$ dominates near the horizon, represents the limit of pure flux advection.

\paragraph{Competition between radial compression and rotational shear for $\xi>0$.}
Once rotation is present, two competing timescales control the field growth: the compression timescale $t_{\rm comp}\sim r/|v_r|$ and the shear timescale $t_{\rm sh}\sim \left|d\Omega/d\ln r\right|^{-1}$. In sub-Keplerian flows ($\xi<1$), centrifugal support is weaker, so $|v_r|$ is larger and $t_{\rm comp}$ is shorter than in the Keplerian case. This preserves strong radial convergence (Figure~\ref{fig:fig2}) and makes the compression-driven growth of $_r B$ highly effective. Conversely, in a Keplerian flow ($\xi=1.0$), differential rotation ($\Omega\propto r^{-3/2}$) is maximal, minimizing $t_{\rm sh}$. Shear efficiently bends poloidal field lines, preferentially enhancing $_\theta B$ (Figure~\ref{fig:fig1}). Our measurements reflect precisely this balance (Figure~\ref{fig:fig3}). This compression–shear equilibrium also establishes favorable conditions for magnetorotational instabilities (MRI) and magnetic reconnection: strongly sheared Keplerian flows provide an ideal environment for the MRI, while compression-dominated sub-Keplerian flows carrying strong radial fields are naturally susceptible to kink-type instabilities near the axis

\paragraph{Role of the sonic transition and inner relativistic regime.}
The transition from the outer quasi-Newtonian domain (subsonic, $v\sim r^{-2}$) to the inner relativistic domain (supersonic, $v\sim r^{-1/2}$) profoundly alters the flow dynamics. This shift reduces the governing timescale, leading to an accelerated evolution of the stationary magnetic zone. The expansion rate changes from $r_{\rm stat}(t)\propto t^{1/3}$ in the outer subsonic region to $r_{\rm stat}(t)\propto t^{2/3}$ in the inner supersonic region. This faster expansion in the relativistic inner flow explains the rapid amplification and reorganization of the near-horizon field observed in Figures~\ref{fig:fig1} and \ref{fig:fig2}. Importantly, this supersonic character persists even for spiral trajectories with substantial azimuthal velocity, demonstrating that rotation alters the geometry of the sonic surface but not the fundamental fact that the flow becomes causally connected to the horizon. This accelerated evolution implies that the black hole engine can establish a magnetically saturated state on a much shorter timescale than predicted by outer flow dynamics alone.

\subsection{Magnetic Field Structure and Astrophysical Implications}\label{subsec:discussion_implications}

The combined effect of compression, shear, and relativistic geometry results in a complex but predictive inner field structure, with significant implications for black hole energetics and observables.

\paragraph{Poloidal flux supply, MAD onset, and jet launching.}
Our results establish that sub-Keplerian flows ($\xi<1$) naturally enhance radial poloidal flux delivery to the innermost regions (Figures~\ref{fig:fig2} and \ref{fig:fig7}). This efficient advection of a strong $_r B$ is the key prerequisite for the formation of a Magnetically Arrested Disk (MAD) state, which maximizes jet power via the Blandford–Znajek mechanism. While the purely non-rotating flow ($\xi=0$) represents the ideal limit of flux accumulation, the sub-Keplerian flow provides the most physically relevant structure that achieves a balance between rotation and strong radial infall. Flows that transiently enter sub-Keplerian (hotter and thicker) states are therefore more likely to accumulate strong near-horizon poloidal fields and launch powerful relativistic jets. This provides a direct physical mechanism for the observed connection between the radiatively inefficient, geometrically thick "hard state" in X-ray binaries (which our sub-Keplerian model resembles) and the presence of powerful, steady jets.

\paragraph{Field geometry, angular momentum transport, and jet coupling.}
The enhancement of $_\theta B$ in Keplerian flows (Figures~\ref{fig:fig1} and \ref{fig:fig4}) indicates stronger shear-driven poloidal bending and the transition to a highly non-radial, shear-dominated structure near the horizon (Figure~\ref{fig:fig6}). This field wrapping is crucial as it generates the magnetic pressure and shear necessary for \textit{angular momentum transport} via magnetic stresses in the disk, consistent with GRMHD findings for SANE disks, where $_\theta B$ provides the kinematic precursor for the dominant $B_\phi$ component. Conversely, the strong $_r B$ component in sub-Keplerian flows suggests a greater vertical magnetic pressure, which can couple the disk and corona, facilitating large-scale magnetocentrifugal acceleration (Blandford–Payne-type) and jet formation. The overall spatial configuration of $_r B$ depletion and $_\theta B$ amplification near the horizon (Figures~\ref{fig:fig4} and \ref{fig:fig5}) confirms the relativistic necessity of an \textit{overwhelmingly non-radial structure} in the energy extraction region.

\paragraph{Comparison with GRMHD and Observational Signatures.}
Our model's findings align well with expectations from complex GRMHD simulations. Our result that reduced rotational support ($\xi<1$) preferentially amplifies $_r B$ provides a direct, physical mechanism explaining the flux advection efficiency seen in thick, hot (ADAF/RIAF-like) states, which favor MAD onset. Conversely, the shear-driven enhancement of $_\theta B$ for $\xi=1$ offers a simplified, physical explanation for the strong shear and field line wrapping prevalent in rotationally supported SANE disks. The faster growth of the stationary magnetic zone in the inner relativistic flow ($r_{\rm stat}\propto t^{2/3}$) implies more rapid variability in horizon-scale polarization and Faraday rotation, which may be observable by EHT polarimetric monitoring. State transitions that alter $\xi$ and the compression-shear balance should produce correlated signatures in jet luminosity, structure, and X-ray Comptonization via magnetic dissipation.

\section{Conclusions}\label{sec:conclusion}

In this work, we developed a semi-analytical, general relativistic framework to study the evolution of a large-scale poloidal magnetic field in accretion flows onto a non-rotating (Schwarzschild) black hole. By parameterizing the degree of rotational support via $\xi$, we systematically explored the interplay between \textbf{radial compression} and \textbf{rotational shearing} in amplifying the magnetic field, focusing on the inner, strong-gravity region where relativistic effects dominate.

\subsection{Summary of Key Findings}
Our primary findings are summarized as follows:
\begin{enumerate}
    \item \textbf{Dominance of Radial Compression:} The purely radial, non-rotating inflow ($\xi=0$) proves to be the most efficient configuration for amplifying both the radial ($B_r$) and polar ($B_\theta$) components of the magnetic field. The absence of centrifugal support allows the plasma to fall rapidly toward the black hole, producing the strongest compression and consequently the highest magnetic amplification. This highlights the significant moderating effect of angular momentum on field growth.
   
    \item \textbf{Kinematic Dichotomy in Rotating Flows:} For flows with rotation ($\xi>0$), amplification mechanisms bifurcate. Sub-Keplerian flows, characterized by slower azimuthal motion and faster radial infall, preferentially amplify the radial component ($B_r$) due to enhanced compression. In contrast, Keplerian flows, with maximal differential rotation, more efficiently amplify the polar component ($B_\theta$) through rotational shearing. This illustrates how the balance between $B_r$-driven advection and $B_\theta$-driven winding governs the resulting magnetic geometry.
   
    \item \textbf{Accelerated Evolution in the Relativistic Regime:} The transition from a quasi-Newtonian, subsonic outer flow to a supersonic, relativistic inner flow leads to accelerated magnetic field evolution. The characteristic growth timescale of the stationary, magnetically saturated region near the black hole is shorter, with expansion scaling as $r_{\rm stat} \propto t^{2/3}$ compared to $r_{\rm stat} \propto t^{1/3}$ in the outer disk \cite{PaperI}.
   
    \item \textbf{Implications for MAD and Jet Launching:} Sub-Keplerian accretion phases (e.g., thick, ADAF-like flows) are kinematically crucial for efficient transport and amplification of radial magnetic flux ($B_r$) toward the horizon. This strong $B_r$ is necessary for MAD onset, confirming the link between low-angular-momentum flows and the launching of relativistic jets.
\end{enumerate}

\subsection{Limitations and Future Work}
While our semi-analytical framework successfully isolates key physical mechanisms, its idealizations define clear directions for future research:
\begin{itemize}
    \item \textbf{Ideal MHD and No Explicit Diffusion:} We impose the frozen-in condition and neglect explicit resistivity. In reality, finite magnetic Prandtl numbers ($P_m \sim \eta/\nu$) and turbulent reconnection can significantly affect compression–shear competition, particularly in thin or moderately thick disks. Including turbulent or physical resistivity would quantify departures from ideal advection-dominated growth.
    
    \item \textbf{Schwarzschild Background:} The black hole spin is set to zero ($a=0$). Extending to Kerr spacetime would incorporate frame-dragging (Lense-Thirring precession) and ISCO shifts, critical for jet power predictions via the Blandford–Znajek mechanism.
    
    \item \textbf{Axisymmetry and Toroidal Field Omission:} Our focus is on the poloidal field; including the toroidal component ($B_\phi$) and non-axisymmetric modes would capture MRI-driven turbulence, dynamo effects, angular momentum transport, and jet collimation, enabling realistic 3D GRMHD comparisons.
    
    \item \textbf{Fixed Flow Kinematics:} The model calculates field evolution on a prescribed velocity field, neglecting back-reaction of magnetic pressure and tension. A self-consistent model solving both flow and fields is required for more realistic dynamics.
    
    \item \textbf{Thermodynamics and Radiation:} Radiative cooling/heating and conduction are omitted. Accounting for these effects would modify disk aspect ratio ($H/R$), sound speed, and compression–shear balance, influencing growth rates of $B_r$ and $B_\theta$, and enabling spectral predictions.
    
    \item \textbf{Boundary Conditions:} While inner solutions are anchored to quasi-Newtonian outer conditions, a fully global solution across the sonic surface would improve quantitative accuracy of the stationary magnetic zone and inner poloidal structure.
\end{itemize}

Addressing these limitations will strengthen physical realism and facilitate direct comparison with GRMHD simulations and observations of M87$^*$ and Sgr A$^*$. Overall, the degree of rotational support emerges as a key parameter controlling both amplification rates and resulting field geometry near the black hole. This work provides a clear physical bridge between inflow kinematics and near-horizon magnetic structure, offering testable predictions for future observational and numerical studies.

\clearpage

\counterwithin{equation}{section}
\begin{appendices}
\section{Analytical Fitting for the Relativistic Inflow Trajectory}\label{sec:appendixA}

The trajectory of a plasma element in the inner, strong-gravity region is described by the integral in Equation~\eqref{eq:magTime}. To obtain a closed-form expression essential for our semi-analytical model, we fit the numerical solution of this integral to a general functional form, previously presented in \cite{PaperI} and reproduced here for clarity:
\begin{equation}
\begin{split}
    \frac{ct}{r_g} &+ a x^{-\frac{5}{2} } + b x^{-\frac{3}{2} } + c x^{-\frac{1}{2} }
    + d \ln \frac{1 - \sqrt{x}}{1 + \sqrt{x}} 
    + f \left( \ln \frac{1 - \sqrt{x}}{1 + \sqrt{x}} \right)^2 \\
    &= a x_0^{-\frac{5}{2} } + b x_0^{-\frac{3}{2} } + c x_0^{-\frac{1}{2} }
    + d \ln \frac{1 - \sqrt{x_0}}{1 + \sqrt{x_0}}
    + f \left( \ln \frac{1 - \sqrt{x_0}}{1 + \sqrt{x_0}} \right)^2.
\end{split}
\label{eq:A1_fitting_function}
\end{equation}
The fitting coefficients ($a, b, c, d, f$) are determined numerically and depend on the flow's angular momentum parameter, $\xi$.

\noindent A crucial distinction arises when analyzing the dominant terms of this function in the relativistic regime studied here, as opposed to the quasi-Newtonian domain of \cite{PaperI}. In the outer regions ($x \to 0$), power-law terms such as $x^{-3/2}$ governed the dynamics. However, in the innermost region, as a fluid element approaches the event horizon ($x \to 1$), the behavior of the trajectory changes fundamentally.

\noindent In this relativistic limit, the logarithmic term, $\ln((1 - \sqrt{x})/(1 + \sqrt{x}))$, becomes the dominant component of the solution. This term diverges as $x \to 1$, accurately capturing the infinite stretching of spacetime as viewed by a distant observer. Our numerical fitting procedure confirms this physical expectation: the coefficient $d$ associated with the primary logarithmic term is found to be the largest, making its contribution decisive in describing the plasma's final plunge into the black hole.

\noindent This shift in the dominant term from a power-law in the Newtonian regime to a logarithmic function in the relativistic regime is not merely a mathematical nuance; it is a direct reflection of the underlying change in physics. It signifies the transition from a flow governed by quasi-Newtonian gravity to one completely dominated by the extreme spacetime curvature near the event horizon. The asymptotic solution for $x_0(x,t)$ presented in Equation~\eqref{eq:x0_asymptotic} is derived from this logarithmic dominance and provides the foundation for our analytical solution of the magnetic field evolution in this strong-field environment.

\end{appendices}

\end{document}